\documentclass[a4paper,fleqn,usenatbib]{mnras}

\usepackage{newtxtext,newtxmath}
\usepackage[T1]{fontenc}
\usepackage{ae,aecompl}

\usepackage{graphicx}	% Including figure files
\usepackage{amsmath}	% Advanced maths commands
\usepackage{amssymb}	% Extra maths symbols

\title[LAE clustering in MUSE-Wide]{The MUSE-Wide survey: Detection of a clustering signal from Lyman$\alpha$-emitters at $3<z<6$}

\author[C. Diener et al.]{
C. Diener$^{1,2}$\thanks{E-mail: cdiener@ast.cam.ac.uk}
L.Wisotzki$^{2}$,
K. B. Schmidt$^{2}$,
E. C. Herenz$^{2,3}$,
T. Urrutia$^{2}$,
T. Garel$^{4}$,
\newauthor
J. Kerutt$^{2}$,
R. L. Saust$^{2}$,
R. Bacon$^{4}$,
S. Cantalupo$^{5}$,
T. Contini$^{6}$,
B. Guiderdoni$^{4}$,
\newauthor
R. A. Marino$^{5}$,
J. Richard$^{4}$,
J. Schaye$^{7}$,
G. Soucail$^{6}$,
P. M. Weilbacher$^{2}$
\\
$^{1}$ Institute of Astronomy, Madingley Road Cambridge, CB3 0HA, UK \\
$^{2}$ Leibniz-Institut f\"ur Astrophysik Potsdam (AIP), An der Sternwarte 16, D-14482 Potsdam, Germany\\
$^{3}$ Department of Astronomy, Stockholm University, AlbaNova University Centre, SE-106 91, Stockholm, Sweden\\
$^{4}$ Univ Lyon, Univ Lyon1, Ens de Lyon, CNRS, Centre de Recherche Astrophysique de Lyon UMR5574, F-69230, Saint-Genis-Laval, France\\
$^{5}$ ETH Zurich, Institute of Astronomy, Wolfgang-Pauli-Str. 27, CH-8093 Zurich, Switzerland\\
$^{6}$ Institut de Recherche en Astrophysique et Plan\'etologie (IRAP), Universit\'e de Toulouse, CNRS, UPS, F-31400 Toulouse, France \\
$^{7}$ Leiden Observatory, Leiden University, P.O. Box 9513, 2300 RA Leiden, The Netherlands \\
}

\date{Accepted XXX. Received YYY; in original form ZZZ}

\pubyear{2017}

\begin{document}
\label{firstpage}
\pagerange{\pageref{firstpage}--\pageref{lastpage}}
\maketitle

\begin{abstract}
We present a clustering analysis of a sample of 238 Ly$\alpha$-emitters at redshift $3\lesssim z\lesssim6$ from the MUSE-Wide survey. This survey mosaics extragalactic legacy fields with 1h MUSE pointings to detect statistically relevant samples of emission line galaxies. We analysed the first year observations from MUSE-Wide making use of the clustering signal in the line-of-sight direction. This method relies on comparing pair-counts at close redshifts for a fixed transverse distance and thus exploits the full potential of the redshift range covered by our sample. A clear clustering signal with a correlation length of $r_{0}=2.9^{+1.0}_{-1.1}$\,Mpc (comoving) is detected. Whilst this result is based on only about a quarter of the full survey size, it already shows the immense potential of MUSE for efficiently observing and studying the clustering of Ly$\alpha$-emitters. 
\end{abstract}


%%%%%%%%%%%%%%%%%%%%%%%%%%%%%%%%%%%%%%%%%%%%%%%%%%

%%%%%%%%%%%%%%%%% BODY OF PAPER %%%%%%%%%%%%%%%%%%

\section{Introduction}
The study of galaxies between redshifts $z\sim 3-6$ is key to our understanding of galaxy formation processes and the evolution from young progenitor galaxies to galaxies in the local Universe. However, accumulating a representative sample of high redshift galaxies is observationally extremely challenging. The most common techniques to reach stastically relevant samples include the search for Lyman-break galaxies (LBGs), exploiting the drop in the continuum bluewards of 912\,$\AA$ (Steidel \& Hamilton 1992), and the observation of Ly$\alpha$-emitters (LAEs) via narrowband (NB) excess (Cowie \& Hu 1998). Both of these techniques are fundamentally photometric approaches with potentially large contamination of the observed samples; in the case of LAEs typical spectroscopic confirmation rates can be as poor as $50\%$ (depending on the combination of NB and broadband filters used for the NB excess selection; e.g. Rhoads et al. 2000, Hu et al. 2010).

Both LBG and LAE samples consist of star-forming galaxies, however with some differences between them resulting from the selection technique (at least to some degree). Compared with typical continuum selected galaxies, LAEs can in principle be probed to much fainter luminosities due to the brightness of the Ly$\alpha$ emission line, even if there may be significant overlap between LBG and LAE surveys (see Yuma et al. 2010 for a more extensive discussion). Whilst both selection techniques have their strengths, this study will focus on a sample of LAEs using the unique capabilities of the  VLT Multi Unit Spectroscopic explorer (MUSE).

Clustering analyses of LAEs are particularly valuable to understand which subpopulation of galaxies they represent at their observed redshift. This helps in furthering our understanding of the impact of environment onto LAE visibility. In combination with cosmological simulations it allows us to link LAEs to their descendants at lower redshift and hence give insight into the evolution of Ly$\alpha$ emitting galaxies. There have been significant efforts at various redshifts, mostly relying on NB selected samples, occasionally with some spectroscopic follow up. At $z=3-4$, Gawiser et al. (2007) and Bielby et al. (2016) analysed 162 and 600 LAEs respectively, finding a correlation length of $r_{0}=3.5-4$\,Mpc and concluding that LAEs typically occupy dark matter haloes with masses of $\sim10^{11}$\,M$_{\odot}$. Furthermore Gawiser et al. (2007) predict that $z\sim3$ LAEs evolve into $\sim L_{*}$ galaxies by $z=0$, although this result depends on the flux limit of their NB selected sample ($1.5\times10^{-17}$\,erg\,s$^{-1}$\,cm$^{-2}$) and may not be true for fainter flux limits. At $z>3$ LAEs probe more and more dense regions of the Universe and LAEs at $z=4-5$ probe galaxies which evolve into $>2.5L_{*}$ objects by the present day.
 At slightly lower redshift, $z=2.1$, Guaita et al. (2010) studied a sample of 250 LAEs, measuring a correlation length $r_{0}=4.8$\,Mpc and predicting $L_{*}$-type descendants, similar to Gawiser et al. (2007).

At $z\sim4.5-5$ Kova{\v c} et al. (2007), Ouchi et al. (2003), Shimasaku et al. (2004) and Shioya et al. (2009) measured correlation lengths of $r_{0}=4.5-5$\,Mpc, whilst Shimasaku et al. (2004) point out large cosmic variance on scales of $\sim70$\,Mpc. Ouchi et al. (2010) measure $r_{0}=3-7$\,Mpc for their $z=6.6$ sample. Most recently Ouchi et al. (2017) measured $r_{0}=4.3$\,Mpc at $z=5.7$ and $r_{0}=3.8$\,Mpc at $z=6.6$. The increase of $r_{0}$ with redshift (even if noisy, see also Figure \ref{fig_lit}) indicates that LAEs occupy denser regions of the Universe at higher redshift. This results in an increased clustering strength and hints towards a constant host halo mass of $\sim10^{11}$\,M$_{\odot}$.

All of the above mentioned studies relied on narrow-band excess selected samples of LAEs and are therefore limited to a single redshift slice; the one covered by the narrowband filter. Consequently it is only possible to infer clustering measurements at one specific redshift. This allows to pinpoint the clustering length at that redshift, however any redshift evolution can only be studied through the combination of multiple surveys. With the advent of the MUSE instrument (Bacon et al. 2010) it has become possible to efficiently sample representative areas of the sky and obtain spectroscopic information without suffering from the poor contrast, uncertain photometric redshifts and sampling rates below 100\% for spectroscopic follow-ups that are typical for LAE samples. Even more importantly, a MUSE survey samples the whole redshift range accessible to the instrument's spectral range, allowing for a LAE sample within a contiguous area and with a continuous redshift distribution, similar to spectroscopic surveys at lower redshifts.

This paper describes the first analysis of LAE clustering in the MUSE-Wide survey, designed to detect LAEs at $3\lesssim z\lesssim6$ over $\sim100$\,arcmin$^2$ at completion. The aim of this paper is to present the analysis of the first quarter of data available from MUSE-Wide and lay the groundwork for the more detailed analysis that will be possible with the complete sample of MUSE-Wide LAEs. We use a method developed by Adelberger et al. (2005) that relies on clustering in the radial ($z$) direction rather than the popular angular clustering method and show its applicability and strengths for use with spectroscopic surveys.
 
 Where applicable we use a $\Lambda$CDM concordance cosmology and adopt $\Omega_{m}=0.3 , \Omega_{\Lambda}=0.7$ and $h=0.7$. Comoving distances are denoted by a leading "c", so comoving Megaparsec becomes "cMpc".

\begin{figure}
\begin{center}
\includegraphics[scale=0.45]{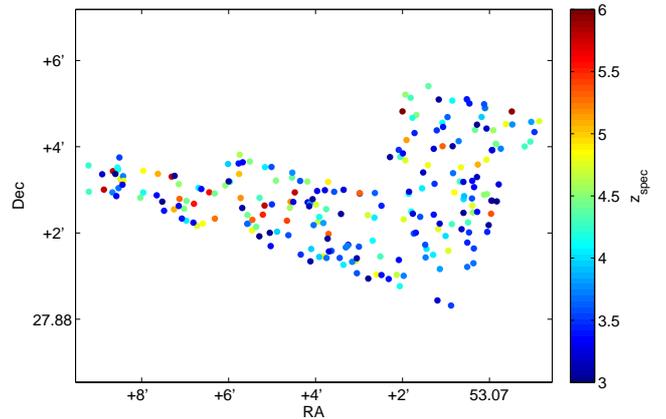}
\end{center}
\caption{Positions of the 238 LAEs in our sample. The individual objects are colour-coded according to their Ly$\alpha$ redshift. The field consists of 24 individual $1'\times1'$ MUSE pointings.}
\label{fig_radec}
\end{figure}

\section{Data}
\subsection{The MUSE-Wide survey}
The MUSE-Wide survey (PI: L. Wisotzki, Programme-ID 094.A-0205(B)) aims at observing a statistically relevant number of emission-line objects in extragalactic legacy fields (Chandra Deep Field South and COSMOS) with pre-existing deep HST data. This allows to complement the spectroscopic information obtained with MUSE with already existing multiwavelength observations which can be used to assess the physical properties of the observed objects such as their stellar mass and star formation rate. Upon completion the survey will have observed $\sim100$ $1'\times1'$ fields with 1h of exposure time each, and result in the detection of $\sim1000$ Ly$\alpha$-emitters. In addition to the primary goal of observing a large LAE sample, MUSE-Wide finds an even greater number of intermediate redshift objects, mostly through the [OII], [OIII] and H$\alpha$ lines. These allow a number of interesting studies themselves, especially in combination with the multiwavelength data.

This paper will focus on the first 24 fields of the MUSE-Wide survey which have been observed within the context of the first year of the guaranteed time observations allocated to the MUSE consortium. All 24 fields lie within the Chandra Deep Field South and have been observed with MUSE with 1h integration time to cover the spectral range from $4750-9350$\,\AA (corresponding to a Ly$\alpha$ redshift range of $2.9-6.7$). Observations took place under clear to photometric conditions and at a mean auto-guider seeing of 0.89'' (for more details see Herenz et al. 2017). Due to slight overlaps of the individual fields the total survey area from these 24 fields amounts to 22.2\,arcmin$^2$. 

The data reduction has been performed with the standard MUSE pipeline to produce fully calibrated data cubes (Weilbacher et al. 2006, 2012 and 2014) and augmented with the ZAP algorithm (Soto et al. 2016) as well as custom-made procedures for better sky subtraction. The details of the reduction process will be discussed in Urrutia et al. (in prep).

\subsection{The Ly$\alpha$-emitter sample}
A detailed description of the emission line source catalogue resulting from the first 24 fields of the MUSE-Wide survey can be found in the dedicated paper Herenz et al. (2017). Here we will only give an overview of the procedure. As mentioned above, the primary goal of the MUSE-Wide survey is to build a sample of emission-line objects. To detect these sources the dedicated software LSDCat\footnote{\url{http://ascl.net/1612.002}} (Herenz \& Wisotzki 2017) was developed and applied to the reduced and flux-calibrated MUSE data cubes. LSDCat uses a matched filter approach assuming a 3D Gaussian profile for detecting emission lines in the full 3D $(x,y,\lambda)$ MUSE data cube and using a detection threshold of $S/N>8$. Prior to an LSDCat detection run continuum flux was removed from the data cubes by applying a median filter in spectral direction.

The candidate emission-line objects resulting from the LSDCat detection have been classified by three independent inspectors. A quality and confidence flag was assigned to each detection. The catalogue from the first 24 MUSE-Wide fields contains 831 emission-line galaxies out of which 238 are LAEs at $z\gtrsim3$. We show their positions in Figure \ref{fig_radec}. One object in our sample was actually detected through its CIV line, whilst the Ly$\alpha$ line was still visible, and was hence included in the analysis here. Most of the LAEs had only one line detected above the signal-to-noise detection threshold (except for 2 AGNs and another object where CIV was found), and have been assigned varying confidence flags, ranging from 1 (uncertain) to 3 (very certain), depending on how clearly the line was identified to actually be Ly$\alpha$. For the overwhelming majority of the objects (218) there is no or only little doubt that they are indeed LAEs (confidence 2 or 3). For the remaining 20 objects there remained substantial doubt on the correct line identification, for instance due to atypcial line-profiles or low S/N). We nevertheless included them in our analysis but verified that their exclusion does not change our results in a significant way. Furthemore we also included the 2 AGNs into the sample again checking that excluding them would not alter our result.

\begin{figure}
\begin{center}
\includegraphics[scale=0.45]{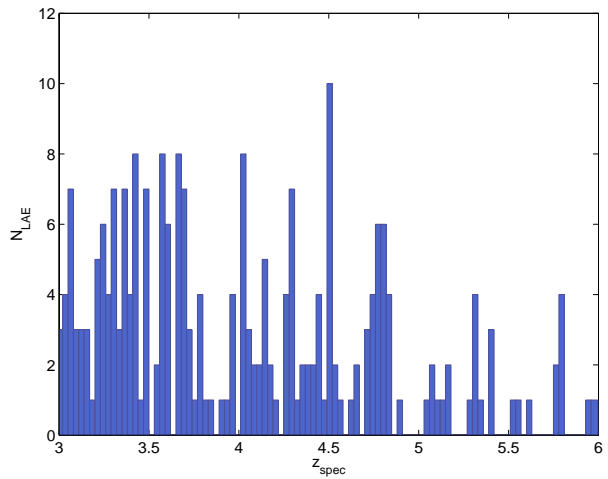}
\end{center}
\caption{Redshift distribution for the 238 LAEs from the MUSE-Wide survey, with a bin-size of $\Delta z = 0.03$, which approximately corresponds to the binning used for the clustering analysis. The redshifts have been measured by fitting an asymmetric Gaussian profile to the reddest peak of the Ly$\alpha$ emission-line.}
\label{fig_nz}
\end{figure}

The redshifts of the LAEs were estimated by fitting an asymmetric Gaussian profile to the Ly$\alpha$ emission line, where in case of double peaks only the red peak was fitted. The resulting redshift distribution is shown in Figure \ref{fig_nz}. The LAE sample has a mean redshift of $\langle z\rangle=4.02$ and a mean redshift error of 0.00051, corresponding to a $\approx30$\,km/s velocity error. We will be using these Ly$\alpha$ redshifts for our analysis as there are typically no other lines available from MUSE-Wide to measure the systemic redhifts.
Whilst it is well known that redshifts estimated from the Ly$\alpha$-line exhibit offsets of roughly a couple of hundred km/s with respect to the systemic redshifts (McLinden et al. 2011, Rakic et al. 2011, Hashimoto et al. 2015, Trainor et al. 2015), for the purpose of the analysis of LAE clustering it is only important that the redshifts are measured self consistently as was done here. Furthermore, errors of a couple of hundred km/s correspond to positional uncertainties of $\sim2-3$ cMpc (radially) and the method employed in this work is quite insensitive to redshift errors of this magnitude (we will be working on radial scales of 25-50\,cMpc). Despite the additional uncertainty due to the use of the non-systemic redshift, this spectroscopic approach delivers an order of magnitude more accurate redshifts than typically obtained by a pure photometric approach, where filter-widths translate to uncertainties of a few 1000\,km/s.

The Ly$\alpha$ flux of the emitters in MUSE-Wide is typically $\sim 1-3\times10^{-17}$\,erg\,s$^{-1}$\,cm$^{-2}$, with an actual flux cut at $\sim 5\times10^{-18}$\,erg\,s$^{-1}$\,cm$^{-2}$. This compares to typical NB studies with flux limits of $\sim 1\times10^{-17}$\,erg\,s$^{-1}$\,cm$^{-2}$. The mean equivalent width has been estimated as 115\,\AA with most emitters lying in the range $37-201$\,\AA (10th and 90th percentile). Again this is similar to previously conducted NB studies where the EW limit ist typically of order $\sim 80$\,\AA. It should however be stressed that MUSE-Wide is a flux-limited survey with the flux limit depending on wavelength.

 For the LAEs in our sample that have a counterpart in deep HST surveys (172 objects) the median stellar mass is $10^{8.7}$\,M$_{\odot}$, as estimated from template fitting with FAST (Kriek et al. 2009)\footnote{Using 3D-HST photometry, the Bruzual\,\&\,Charlot (2003) stellar library and a truncated constant star formation history.} This compares to other LAE surveys at $z\sim3$ which typically consist of low mass objects ($\sim 10^9$\,M$_{\odot}$) which are actively star-forming with star formation rates of a few solar masses per year (e.g. Gawiser et al. 2007, Ono et al. 2010).

\section{Methods}
Typical clustering analyses of LAEs are conducted by using a single redshift slice and thus limited to estimating the clustering length at only that specific redshift. Usually the observed angular clustering is measured and the angular correlation function is then related to the three-dimensional correlation function via the so-called Limber equation (Limber 1953). It exploits the fact that the observed clustering is essentially just a projection of the three-dimensional clustering and can be deprojected if the redshift distribution function is known accurately. In the case of spectroscopic surveys this technique is typically limited by non-random slit allocation causing artificial clustering and by only observing a small fraction of the objects. For photometric surveys the redshift distribution is often uncertain resulting in large uncertainties in the deprojection. 

With a 22.2\,arcmin$^2$ survey area, the first 24 MUSE-Wide pointings studied here cover a smaller field than previous LAE studies, but span a large continuous redshift range of $3\lesssim z\lesssim6$. It is therefore a logical step to use a method that is virtually orthogonal to the standard angular clustering approach, by exploiting the clustering in the redshift direction instead. Such a method was first introduced by Adelberger et al. 2005 (A05 hereafter) and essentially relies on pair-counting at close redshifts. A05 applied it to a sample of $1.4\lesssim z\lesssim3.5$ star-forming galaxies and showed that it yields results consistent with the angular clustering method.

The basic idea of A05 is to compare the number of galaxy pairs $N$ at fixed transverse distance $R_{\textup{ij}}$ and different radial distances $Z_{\textup{ij}}$ from each other. Assuming bins $a_{1}<Z_{\textup{ij}}<a_{2}$ and $b_{1}<Z_{\textup{ij}}<b_{2}$, the estimator adopted is defined as:
 $$K_{a_{1},a_{2}}^{b_{1},b_{2}}(R_{\textup{ij}})\equiv \frac{ N_{b_{1},b_{2}}(R_{\textup{ij}})}{N_{b_{1},b_{2}}(R_{\textup{ij}})+N_{a_{1},a_{2}}(R_{\textup{ij}})},$$
where $N_{b_{1},b_{2}}$ and $N_{a_{1},a_{2}}$ denote the number of galaxies in the respective bins. In the case of no clustering and equal bin sizes this would reduce to $K_{a_{1},a_{2}}^{b_{1},b_{2}}=0.5$ for all transverse distances.

Following A05 the expectation value of this estimator can be calculated from the three-dimensional correlation function $\xi(r)$ as follows:
\begin{multline*}
<K_{a_{1},a_{2}}^{b_{1},b_{2}}(R_{ij})> \simeq \Big{\{}(b_{2}-b_{1})\sum\limits_{\scriptscriptstyle i>j}^{\scriptscriptstyle \textup{pairs}}[1+\bar{\xi}_{b_{1},{b_{2}}}]\Big{\}} \\
\times \Big{\{}(b_{2}-b_{1})\sum\limits_{\scriptscriptstyle i>j}^{\scriptscriptstyle \textrm{pairs}}[1+\bar{\xi}_{b_{1},{b_{2}}}]+(a_{2}-a_{1})\sum\limits_{\scriptscriptstyle i>j}^{\scriptscriptstyle \textrm{pairs}}[1+\bar{\xi}_{a_{1},{a_{2}}}]\Big{\}}^{-1},
\end{multline*}\\
where $\bar{\xi}$ is defined as $$\bar{\xi}_{a_{1},a_{2}}(R_{ij}) = \frac{1}{a_2-a_1} \int_{a_{1}}^{a_{2}} dZ\, \xi(R_{ij},Z).$$

The 3D correlation function $\xi(r)$ has traditionally been found to take the form $\xi(r) = (r/r_{0})^{-\gamma}$. The above expectation value can consequently be used to determine the correlation length $r_{0}$ and exponent $\gamma$ from an observed signal by estimating $K_{a_{1},a_{2}}^{b_{1},b_{2}}$ at various transverse distances $R_{\textup{ij}}$. Often the limited number of objects in a sample will however not permit this simultaneous constraint, in which case $r_{0}$ can still be estimated by assuming a fixed value for $\gamma$ and minimising the distance between the measured $K_{a_{1},a_{2}}^{b_{1},b_{2}}$ and the expectation value. With currently only 238 LAEs observed, we will take this approach. A simultaneous constraint should be possible once the full sample of 1000 LAEs from MUSE-Wide is available.

\section{Results}
\subsection{Measuring the clustering signal}
As outlined in the previous section, we are relying on a method developed by A05 to estimate the LAE clustering solely from the clustering in radial direction by counting LAE pairs with close redshifts. To estimate the clustering in our sample we adopted the radial separations of $0<Z_{ij}<25$\,cMpc and $25<Z_{ij}<50$\,cMpc (see also A05) and calculated the respective numbers of galaxy pairs at given transverse distances $R_{\textup{ij}}$. Clearly, too large bins would make $K_{a_{1},a_{2}}^{b_{1},b_{2}}$ insensitive to a possible clustering signal: we found that values from $(a_{2},b_{2})=(35,70)$\,cMpc and above result in a clear drop and eventual disappearance of the signal.
 On the other hand too low values would i) reduce the numbers of pairs significantly and hence increase the error bars considerably and ii) at some point only probe small-scale clustering. For the MUSE-Wide sample this is the case for values of $(15,30)$\,cMpc and below. In between those extremes, the signal however converges and we have verified that our result, and in particular our estimate of $r_{0}$ (see next section), does not depend critically on the exact values of $a_{1}$, $a_{2}$, $b_{1}$ and $b_{2}$. 

\begin{figure}
\begin{center}
\includegraphics[scale=0.45]{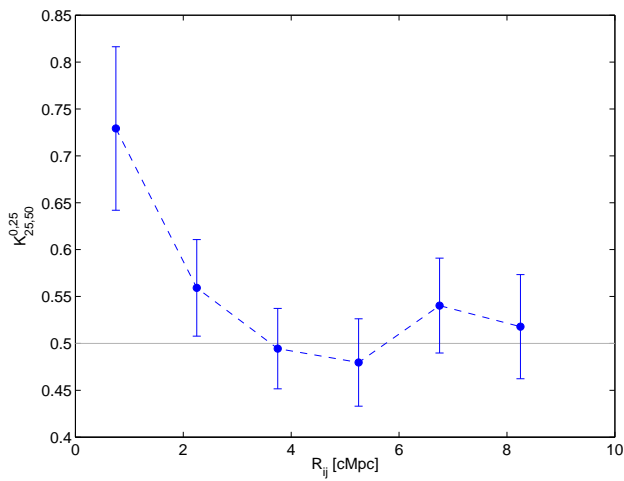}
\end{center}
\caption{The estimator $K_{25,50}^{0,25}$ as a function of comoving transverse pair distance, exhibiting a clear clustering signal out to $\approx3$\,cMpc. The grey line marks the expectation for no clustering. As described in the text, $K_{25,50}^{0,25}$ is essentially the ratio of LAE pairs within 25\,cMpc line-of-sight distance versus pairs within 50\,cMpc line-of-sight distance from each other. The error bars have been calculated as $\sqrt n/d$ where $n$ and $d$ are the numerator and denominator of $K_{25,50}^{0,25}$.}
\label{fig_corr}
\end{figure}

The result of estimating $K_{25,50}^{0,25}$ as a function of comoving transverse pair distance is shown in Figure \ref{fig_corr} with error bars calculated as $\sqrt n/d$ where $n$ and $d$ are the nominator and denominator of $K_{25,50}^{0,25}$. We clearly detect a positive clustering signal, in particular at $R_{ij}\lesssim3$\,cMpc (corresponding to $\sim1.4$\,arcmin at $z=4$). The limited scale of the MUSE-Wide survey prohibits us to measure the signal at larger transverse distances.

Since we are measuring the clustering signal from the redshifts of the Ly$\alpha$-emitters, the impact of any uncertainties in the redshift measurements may be a concern. We investigated the impact of redshift errors by perturbing each redshift by an error drawn from a Gaussian with a standard deviation equaling the mean of the redshift error in the survey ($\Delta z = 0.00051$). With these perturbed redshifts the estimator $K_{25,50}^{0,25}$ was re-calculated. When repeating the procedure 100 times we find that the uncertainties introduced by the redshift errors are much smaller than the uncertainties arising from the limited number of sources. Given the small redshift errors in the MUSE-Wide survey ($\approx30-40$\,km/s) this result is not unexpected.

As discussed already, the redshifts in the MUSE-Wide sample are measured from the Ly$\alpha$ line which is offset from the systemic redshift of the source. With a constant offset this would have no impact onto our analysis, however the spread in redshift offsets in principle acts as an additional redshift uncertainty. As outlined in Hashimoto et al. (2015) this spread is of order of the redshift error in the MUSE survey, meaning that the analysis on the impact of redshift uncertainties as described above is valid in this case as well.

We also verified that our clustering signal is not dominated by the one-halo term by excluding all pairs with transverse separations up to $R_{ij}=0.1$\,Mpc (physical). This corresponds to about twice the typical virial radius of  haloes hosting $\sim10^9$\,M$_{\odot}$ galaxies at $z\sim4$ as derived from the Millennium simulation (Springel et al. 2005).

As the MUSE-Wide survey spans a large redshift range, it may be argued that the higher redshift objects have to be intrinsically more luminous objects to be detected and may therefore dominate our clustering signal. To test this scenario we limited the analysis by excluding any Ly$\alpha$-emitter at varying redshift cuts $z>z_{cut}$ ($z_{cut}=4.5-5.5$ and then recalculated the estimator  $K_{25,50}^{0,25}$. It turns out to be virtually indistinguishable from the value of  $K_{25,50}^{0,25}$ when including all objects.

\subsection{Estimating the correlation length $r_{0}$}
As already discussed, the sample size of our current survey does not permit a simultaneous constraint on the correlation length $r_{0}$ and exponent $\gamma$. However, the correlation length can be estimated by calculating the expectation value $<K_{25,50}^{0,25}>$ whilst stacking all pairs up to a transverse distance $R_\mathrm{cut}$ and keeping $\gamma$ fixed. 

We assume the two-point correlation function to take the standard form $\xi(r) = (r/r_{0})^{-\gamma}$ and set $\gamma$ fixed to the canonical $\gamma=1.8$, e.g. Zehavi et al. (2002). Whilst this value is usually assumed if $r_{0}$ and $\gamma$ cannot be constrained simulataneously, it may be an overestimate of the true $\gamma$ (see for example Moustakas \& Somerville 2002 for the redshift dependency of $\gamma$, or Quadri et al. 2007. But note also that Shioya et al. 2009 constrain $\gamma=1.9\pm0.22$ for a sample of $z=4.86$ LAEs). However, our estimate of $r_{0}$ is insensitive to the exact value assumed; varying the value of $\gamma$ from 1.6 to 2.0 at fixed $R_\mathrm{\textup{cut}}$ only changes the result for $r_{0}$  by 3.5\%, which is well within our error bars due to the small sample size.

Including all pairs at transverse distance  $R_\mathrm{\textup{cut}}<5$\,cMpc we found a correlation length of $r_{0}=2.9^{+1.0}_{-1.1}$\,cMpc. Varying $R_\mathrm{\textup{cut}}$ from 3.5\,cMpc to 5.5\,cMpc results in less than 10\% changes in the value for $r_{0}$ and less than 20\% for values higher than $R_\mathrm{\textup{cut}}=6$\,cMpc. At $R_\mathrm{\textup{cut}}\leq3$\,cMpc the values for $r_{0}$ start to depend sensitively on the exact value of $R_\mathrm{\textup{cut}}$ as we are in the steeply rising regime of the $K_{25,50}^{0,25}$-curve.

At the cost of enlarging the errorbars on $r_{0}$, we also estimated its value for two redshift bins, splitting the sample at the median redshift ($z_{\textrm{median}} = 3.88$). Again we used $R_\mathrm{\textup{cut}}<5$\,cMpc and a fixed $\gamma=1.8$. The resulting values are $r_{0}^{\textrm{low}}=1.8^{+1.4}_{-1.8}$\,cMpc for the lower bin and $r_{0}^{\textrm{high}}=4.4^{+1.6}_{-1.6}$\,cMpc at higher redshifts.

\subsection{Comparison to simulations}

In order to understand how our result compares to expectations from dark matter simulations, we performed an analysis by using the mock catalogues presented by Garel et al. (2016).
They used the GADGET code (Springel et al. 2005) to provide the underlying DM framework which was populated with galaxies through semi-analytic modelling. The DM simulation has been run with a box size of $100^3$\,(cMpc/h)$^3$ and WMAP-5 cosmological parameters ($H_{0} = 70$\,km\,s$^{-1}$\,Mpc$^{-1}$, $\Omega_{m} = 0.28$, $\Omega_{\Lambda}= 0.72$ and $\sigma_{8} = 0.82$). The achieved DM halo mass resolution is $M_{\textrm{halo}} = 2\times10^{9}$\,M$_{\odot}$, which corresponds to a $\approx 2\times10^{-19}$\,erg\,s$^{-1}$\,cm$^{-2}$ Ly$\alpha$ flux resolution limit (Garel et al. 2016).
The details of the semi-analytic model can be found in Garel et al. (2015). It has been calibrated to reproduce the LAE and LBG luminosity functions at $3\lesssim z\lesssim7$, the redshift range relevant for LAE studies with MUSE. 

The final output of the simulations has been translated to mock light-cones as described in Garel et al. (2016) to produce observable quantities such as redshift, positions and Ly$\alpha$ fluxes. We use 100 light-cones for our study with a field-of-view of 100\,arcmin$^2$ each. This field-of-view also corresponds to the final survey area of MUSE-Wide.

In order for the mock catalogues to resemble the MUSE-Wide survey, we imposed a cut in Ly$\alpha$ flux, typically of order $1\times10^{-17}$\,erg\,s$^{-1}$\,cm$^{-2}$, to match the number density of MUSE-Wide. We randomly selected a 22.2\,arcmin$^2$ field within the mocks field-of-view and adjusted the redshifts of the mock sample to entail the redshift error of the MUSE-Wide survey by applying a random error drawn from a Gaussian with a standard deviation equaling the mean redshift error of the survey ($\sim40$\,km/s). The mock samples produced closely follow the MUSE-Wide flux distribution. The mean dark matter halo mass of this sample is $\sim 5\times10^{11}$\,M$_{sun}$.

\begin{figure}
\begin{center}
\includegraphics[scale=0.43]{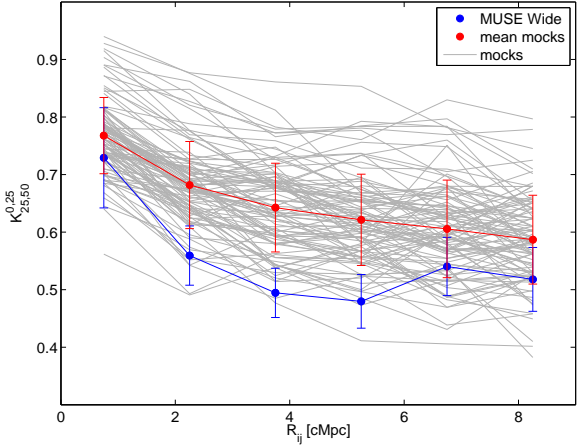}
\end{center}
\caption{Comparison of the clustering signal in the mocks to the MUSE-Wide survey. In grey we show the result from 100 light-cones that replicate MUSE-Wide type observations within the simulations and in red their mean with the standard deviation as error bars. The blue curve stems from the actual data and is the same as in Figure \ref{fig_corr}.}
\label{fig_sim}
\end{figure}

\begin{figure}
\begin{center}
\includegraphics[scale=0.43]{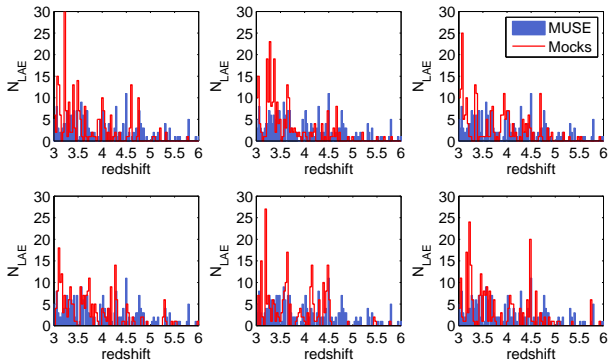}
\end{center}
\caption{Example redshift distributions (red), randomly drawn from the 100 mock lightcones we are using to compare to the data from MUSE Wide (blue). The bin width is $\Delta z = 0.03$, reflecting the binning used in our clustering analysis. The simulated LAEs are much more clustered in redshift than the observed data, also influencing the clustering measurement derived from them.}
\label{fig_simNz}
\end{figure}

From the mock catalogues we calculated $K_{25,50}^{0,25}$. The result is shown in Figure \ref{fig_sim}, where we show both the individual curves (grey) as well as the average over the 100 light cones.
Clearly the clustering in most light-cones is stronger than what is observed with MUSE-Wide. This is most likely due to the fact that the mock observations show large redshift spikes which are not present in the data, as indicated in Figure \ref{fig_simNz} and as tests with modified redshift distributions, exluding those redshift spikes show. These strongly clustered redshift layers lead to dominate the clustering signal. The reason for the appearance of such "super-structures" will need to be assessed by future simulation work.

Finally we used the simulation to check for effects of the survey geometry. This is not expected to play an important role, given that the method relies on radial clustering and is hence devised to lower the effects of geometry. For our test we compared 100 realisations of a square 22.2\,arcmin$^2$ survey versus 100 realisations of the L-shaped type area of MUSE-Wide. We find that the respective clustering signals agree almost perfectly.

\section{Summary and Conclusions}
We have analysed a sample of 238 Ly$\alpha$-emitters observed with the MUSE instrument in a 22.2\,arcmin$^2$ area and spanning the redshift range $3\lesssim z\lesssim6$. This sample arises from the first catalogued objects from the MUSE-Wide survey which, once complete, will observe $\sim1000$ LAEs in about 4$\times$ the current survey area. With its large redshift range, but limited angular coverage, this sample is ideal for applying the Adelberger et al. (2005) method of radial clustering analysis, relying essentially on galaxy pair-counts at close redshifts. 

\begin{figure}
\begin{center}
\includegraphics[scale=0.45]{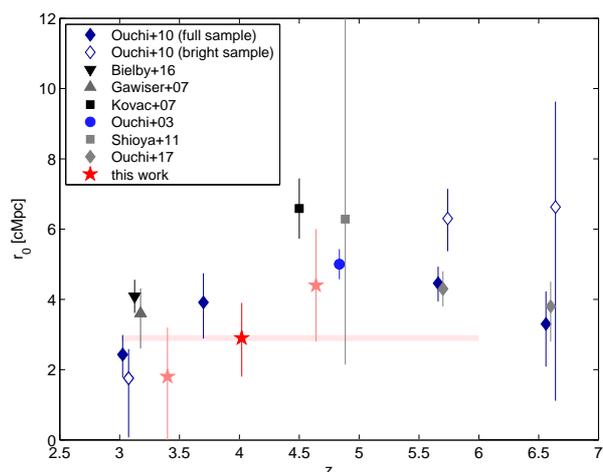}
\end{center}
\caption{Comparison of literature values for $r_{0}$ with the value estimated from MUSE-Wide (red star). Whilst the literature values originate from narrow-band surveys and are therefore restricted to a single redshift slice, the value from our survey comes from the whole redshift range $z=3-6$ (as indicated by the red bar), but is plotted at the mean redshift $<z>=4.02$ of the survey. Also indicated in light red we show the correlation length in two redshift bins splitting the sample at the median redshift.}
\label{fig_lit}
\end{figure}

We found a clear line-of-sight clustering signal at transverse distances up to $\sim3$\,cMpc, and, assuming a correlation function of the form $\xi(r)=(r/r_{0})^{-\gamma}$ with $\gamma=1.8$, we estimate a correlation length of $r_{0}=2.9^{+1.0}_{-1.1}$\,cMpc. Figure \ref{fig_lit} compares this value to previous studies with the same assumed value for $\gamma$ (with the exception of Shioya et al. 2011). Most recently, Bielby et al. (2016) have measured $r_{0}=2.9\pm0.45$\,cMpc for a LAE sample at $z=3.1$. This is similar to Gawiser et al. (2007) with  $r_{0}=3.6^{+0.8}_{-1.0}$\,cMpc at the same redshift. At higher redshifts, $z=4.86$ Ouchi et al. (2003) and Shioya at al. (2009) have found $r_{0}=5.0\pm0.4$\,cMpc and $r_{0}=4.4^{+5.7}_{-2.9}$\,cMpc respectively. Finally Ouchi et. (2010) measured $r_{0}=3-7$\,cMpc for a sample of $z=6.6$ LAEs and refining that Ouchi et al. (2017) estimated $r_{0}=4.3$\,Mpc at $z=5.7$ and $r_{0}=3.8$\,Mpc at $z=6.6$. Most of these surveys have flux limits of order $1-2\times10^{-17}$\,erg\,s$^{-1}$\,cm$^{-2}$, which is a bit higher than for MUSE-Wide. Possibly partly for that reason, the MUSE-Wide estimate of $r_{0}$ is slightly lower than most values reported, however still consistent with previous measurements.
Some of the differences may also be attributed to redshift evolution and the use of differing cosmological values (e.g. assuming $\Omega_{m}=0.25$ instead of $\Omega_{m}=0.3$ would result in $r_{0}=3.5^{+1.0}_{-1.1}$\,cMpc instead of $r_{0}=2.9.5^{+1.0}_{-1.1}$\,cMpc) across the individual surveys.

We also estimated its value for two redshift bins, splitting the sample at the median redshift ($z_{\textrm{median}} = 3.88$). The resulting values are $r_{0}^{\textrm{low}}=1.8^{+1.4}_{-1.8}$\,cMpc for the lower bin and $r_{0}^{\textrm{high}}=4.4^{+1.6}_{-1.6}$\,cMpc in the higher bin. In accordance with the literature we find a higher value for $r_{0}$ at higher redshifts. Again these values are a bit lower than previous estimates from the literature however still within the respective errorbars. We also stress that the combined and individual measurements from the MUSE-Wide data are consistent with each other within the errorbars.

The analysis presented in this paper is, to our knowledge, the first using the redshift pair-count method of Adelberger et al. (2005) applied to a sample of Ly$\alpha$-emitters. Previous studies have relied on the angular clustering method, which is prone to systematic effects due to survey geometry and uncertainties in the redshift distribution. The use of this method was only possible due to the large redshift range and the availability of spectroscopic redshifts from MUSE.
Whilst the current sample still only represents about a quarter of the final survey, we could already demonstrate the detection of a clear clustering signal and the immense potential of the MUSE-Wide survey. With the full sample we expect to be able to look for possible redshift evolution of the clustering signal as well as any dependencies of the clustering strength on LAE properties like star formation rate, mass or equivalent width of the Ly$\alpha$-line.

\section*{Acknowledgements}
This research is based on observations collected at the European Organisation for Astronomical Research in the Southern Hemisphere under ESO programme 094.A-0205(B).

We acknowledge funding by the Competitive Fund of the Leibniz Association through grants SAW-2013-AIP-4 and SAW-2015-AIP-2.

RB acknowledges support from the ERC advanced grant 339659-MUSICOS. RAM acknowledges support by the Swiss National Science Foundation. SC gratefully acknowledges support from Swiss National Science Foundation grant PP00P2\_163824.
This work was supported by the Netherlands Organisation for Scientific Research (NWO), through VICI grant 639.043.409, and the European Research Council under the European Union's Seventh Framework Programme (FP7/2007- 2013) / ERC Grant agreement 278594-GasAroundGalaxies.
TG is grateful to the LABEX Lyon Institute of Origins (ANR-10-LABX-0066) of the Universit\'e de Lyon for its financial support within the programme `Investissements d'Avenir' (ANR-11-IDEX-0007) of the French government operated by the National Research Agency (ANR). 

The Millennium Simulation databases used in this paper and the web application providing online access to them were constructed as part of the activities of the German Astrophysical Virtual Observatory (GAVO).

%%%%%%%%%%%%%%%%%%%%%%%%%%%%%%%%%%%%%%%%%%%%%%%%%%



%\bibliographystyle{mnras}


%%%%%%%%%%%%%%%%%%%%%%%%%%%%%%%%%%%%%%%%%%%%%%%%%%

%%%%%%%%%%%%%%%%% APPENDICES %%%%%%%%%%%%%%%%%%%%%

%%%%%%%%%%%%%%%%%%%%%%%%%%%%%%%%%%%%%%%%%%%%%%%%%%

% Don't change these lines
\bsp	% typesetting comment
\label{lastpage}
\end{document}